\pdfoutput=1
\documentclass[aps,pre,twocolumn]{revtex4} 
\usepackage{hyperref,natbib,doi}
\usepackage{graphicx}
\usepackage{amsmath}
\usepackage{dcolumn}
\usepackage[utf8]{inputenc}
\usepackage{amssymb}
\begin{document}
\title{New regime of plasma wake field acceleration in the extreme blowout regime}
\author{David Tsiklauri}
\affiliation{School of Physics and Astronomy, Queen Mary University of London, London, E1 4NS, United Kingdom}
\begin{abstract}
Three dimensional particle in cell simulations are used for studying 
proton driven plasma wake-field
acceleration that uses a high-energy proton
bunch to drive a plasma wake-field for electron beam acceleration.
A new parameter regime was found
which generates essentially constant electric field that is
three orders magnitudes larger than that of AWAKE design, i.e.
of the order of $2 \times 10^{3}$ GV/m.
This is achieved in the the extreme blowout regime, when number density
of the driving proton bunch exceeds plasma electron number density 100 times.
\end{abstract}

\maketitle

AWAKE is a new particle acceleration experiment 
 currently being built at CERN \citep{awake}. 
The aim of AWAKE is to find a new design for future generation 
high energy particle accelerators. 
This is the first proton driven plasma wake-field
acceleration experiment, which will use a high-energy proton
bunch to drive a plasma wake-field for electron beam acceleration.
A 400 GeV/c proton beam will be extracted from the CERN Super
Proton Synchrotron, SPS, and utilized as a drive beam for wake
fields in a 10 m long plasma cell to accelerate electrons with
amplitudes up to the GV/m level \citep{awake}. 
The plasma acceleration 
based on laser wake field acceleration 
originates from a paper by Tajima and Dawson \citep{td79}. 
The two known possibilities for creation of the plasma wake are:
a laser or an electron bunch. 
Accordingly, they are known as laser wake field acceleration (LWFA)
and the latter as plasma wake field acceleration (PWFA).
A good progress in PWFA has been made 
both in experiment and 
theory \cite{kats83,chen85,kats86,l14,lau15,farinella16,me2016,bera16}. 
Further, Ref.\cite{sr1} reviews the complexities of laser-plasma 
interactions to underline the unique and extraordinary possibilities that the laser ion source offers. 
The considered effects in Ref.\cite{sr1} include keV and MeV ion 
generation, nonlinear (ponderomotive) forces, self-focusing, resonances 
and hot electrons, parametric instabilities, double-layer effects, 
and the few ps stochastic pulsation (stuttering).
Ref.\cite{sr2} showed that an 
energy gain of more than 42 GeV can be achieved in a plasma wakefield 
accelerator of 85 cm length, driven by a 42 GeV electron beam at the 
Stanford Linear Accelerator Center (SLAC). It was shown that the 
results are in excellent agreement with the predictions of 
three-dimensional particle-in-cell simulations. It was found 
that most of the beam electrons lose energy to the plasma wave, 
but some electrons in the back of the same beam pulse are accelerated 
with a field of 52 GVm$^{-1}$. This effectively doubles their energy, 
producing the energy gain of the 3-km-long SLAC accelerator in less 
than a metre for a small fraction of the electrons in the injected 
bunch. Ref.\cite{sr2} was an important step towards demonstrating 
the viability of plasma accelerators for high-energy physics applications. 
Ref.\cite{sr3} gives an exciting, general overview of both plasma 
and laser wakefield acceleration.
Ref.\cite{sr4} is a good reference source for the field of 
high-intensity and high-plasma-density laser-plasma interaction. 
It summarizes past advances and opens insights into the future. 
The book covers the essentials from a single particle to dense fluids, 
and from computational physics to applications for fusion energy or 
hadron cancer therapy. Topic such as
Advanced Laser Acceleration of Electrons, Ultra-fast Acceleration 
of Plasma Blocks by the Nonlinear Force, Laser-Driven Fusion with Nanosecond Pulses,
Laser-Driven Fusion Energy with Picosecond Pulses 
for Block Ignition are covered in depth.
It is interesting to note the
idea of so-called beat-wave accelerator, in which a longitudinal 
resonance field produced in a plasma by the beating of 
two laser beams can accelerate electrons by
$\simeq$ 1 GeV m$^{-1}$. However, this and other types 
of laser accelerators involving plasmas 
encounter problems such as instabilities, 
self-focusing, de-tuning and the appearance of 
internal electric fields and double layers, 
which appear to limit the energy to a few MeV. 
To overcome these difficulties, Ref.\cite{sr5} proposed
a laser accelerator system that does not require a 
plasma, in which charged particles are accelerated 
by nonlinear forces towards minima in a field 
created by two collinear laser beams, and the laser 
intensity gradients are moved in phase with the 
accelerated particles by electro-optical modulation 
of the frequency and/or phase of the beams. 
With sufficiently high laser powers, electrons 
from conventional accelerators could be further 
accelerated by as much as 6 GeV m$^{-1}$.
Ref.\cite{sr6} recently suggested a 
  new scheme of plasma block acceleration based 
  upon the interaction between double targets and 
  an ultra-intense linearly polarized laser pulse 
  with intensity $I \simeq 10^{22}$ W cm$^{-2}$. 
  The work was carried out 
  via two-dimensional particle-in-cell simulations. 
  The targets used were composed of a pre-target of low-density 
  aluminum plasma and an overdense main-target of hydrogen 
  plasma. Through intensive parameter optimization, authors of
  Ref.\cite{sr6} have 
  observed highly efficient plasma block accelerations with 
  a monochromatic proton beam peaked at GeVs. 
  They suggest that the mechanism 
  can be attributed to the enhancement 
  of the charge separation field due to the 
  properly selected pre-target.

The main focus of recent research in the PWFA has been creation of
a constant (essentially spatially flat) electric field of the wake.
The significant new contribution of the present work is to
use fully-kinetic, electromagnetic, 3D PIC simulation to
perform scan of the parameter space, e.g.
(i) varying density of the proton beam; (ii) considering
proton bunches both localized and flat in the transverse y- and z- directions 
(here we only show the best results for the flat, essentially, 1D beam);
(iii) altering proton bunches' energy, thus finally
establishing a new parameter regime. 
The new regime generates essentially constant electric field (which results in
monoenergetic electron beam of high quality, with small energy spread)
three orders magnitudes higher than that of AWAKE design, i.e.
of the order of $2 \times 10^{3}$ GV/m.
This is achieved in the extreme blowout regime, 
namely when number density
of the driving proton bunch exceeds plasma electron 
number density 100 times.
The main purpose of this work is to trigger an interest of
PWFA acceleration experimental community to 
repeat this new regime in their {\it experiments} to
prove its existence. Our {\it numerical} experiments 
suggest such new regime is possible.

\section{The model and results}

\begin{figure*}
\includegraphics[width=\textwidth]{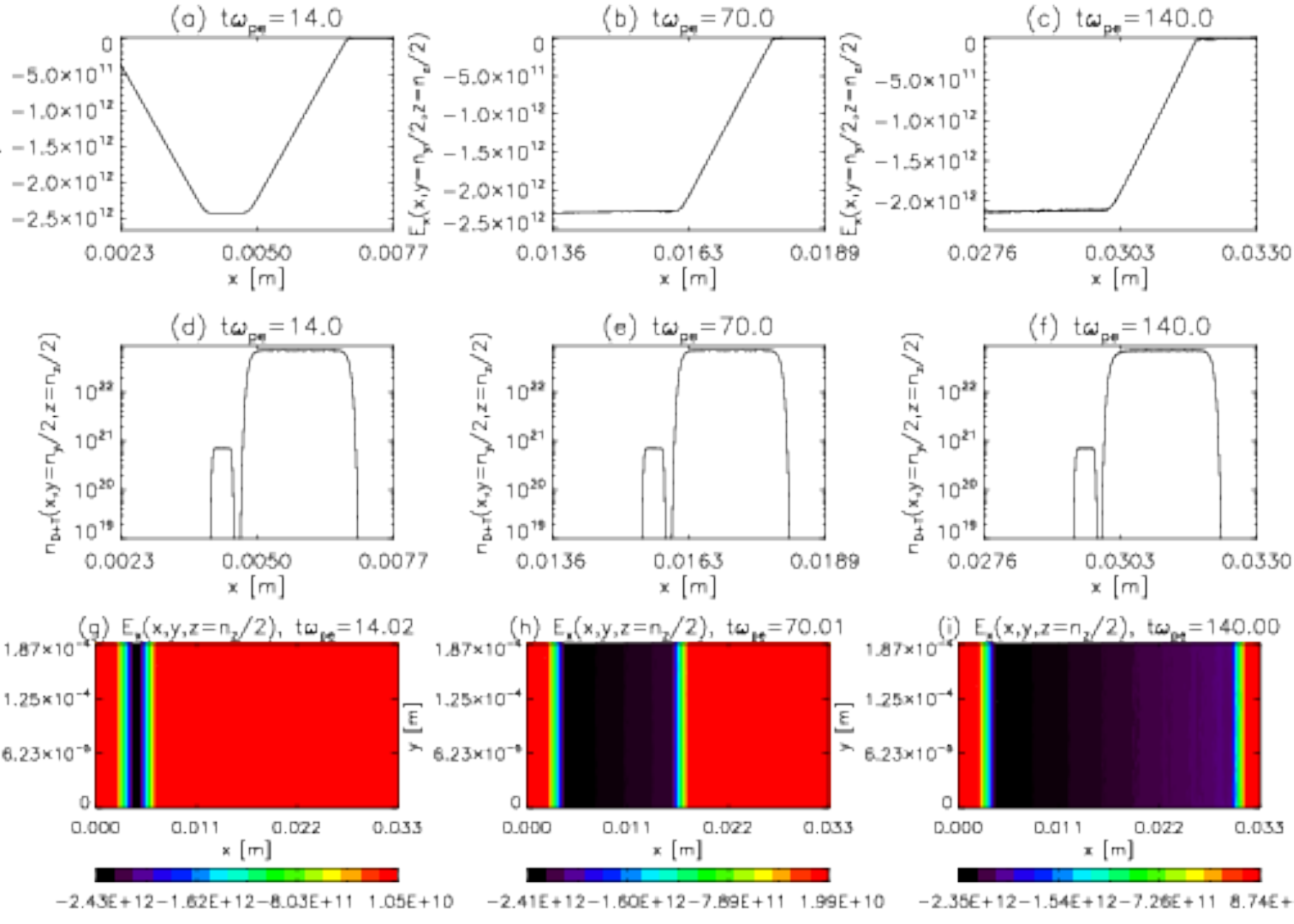}
\caption{(a-c) electric field x-component profile in x-direction, 
$E_x(x,y=n_y/2,z=n_z/2)$, at the middle of the simulation domain in y- and z-,
at different time instants
corresponding to 1/5th, half and the final simulations times.
(d-f) log normal plot of driving proton
 bunch plus electron trailing bunch number densities at the same times.
(g-i) electric field x-component but now for entire simulation domain
 at the same times. 
The fields are quoted in $V/m$ and time at the top of each panel is
in $ \omega_{pe}$. 
Note that x-coordinate is different in panels (a-f) 
because we use a window which follows the 
bunch with speed $v_b=0.9999c$.}
\label{fig1}
\end{figure*}

\begin{figure*}
\includegraphics[width=\textwidth]{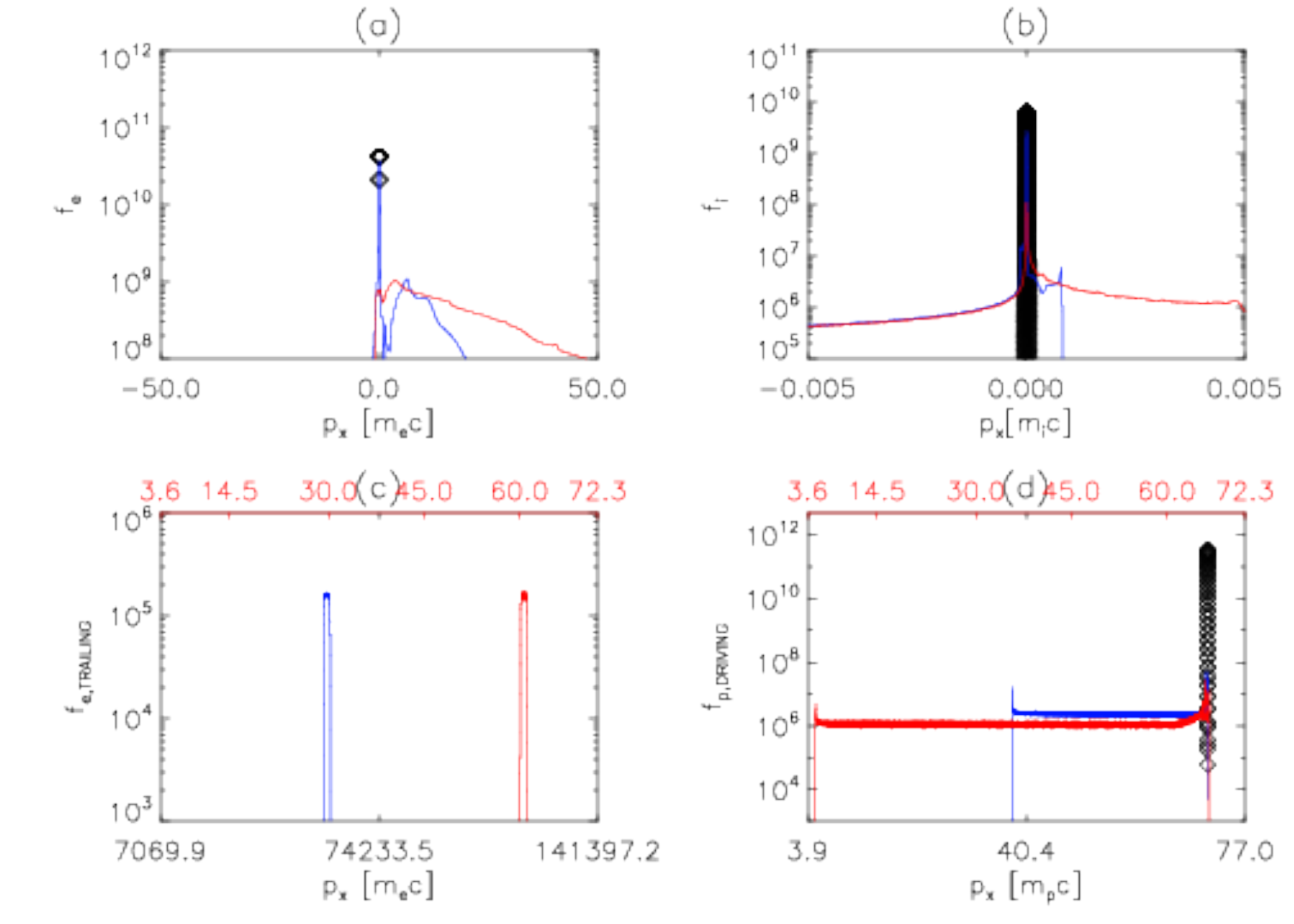}
\caption{Background electron (a), rubidium ion (b), trailing electron bunch (c) and 
driving proton bunch (d) 
distribution functions shown at different times in different colors:
open diamonds correspond to $t=0$, while blue and red curves to the 
half and the final simulations times, respectively. On
x-axis the momenta are quoted in the units of relevant species mass times
speed of light i.e. $[m_e c]$, $[m_i c]$ or $[m_p c]$ as appropriate. 
At the top of panels (c) and (d), to guide the eye,  the energies are stated in GeV with red numbers.}
\label{fig2}
\end{figure*}

\begin{figure} 
\includegraphics[width=\columnwidth]{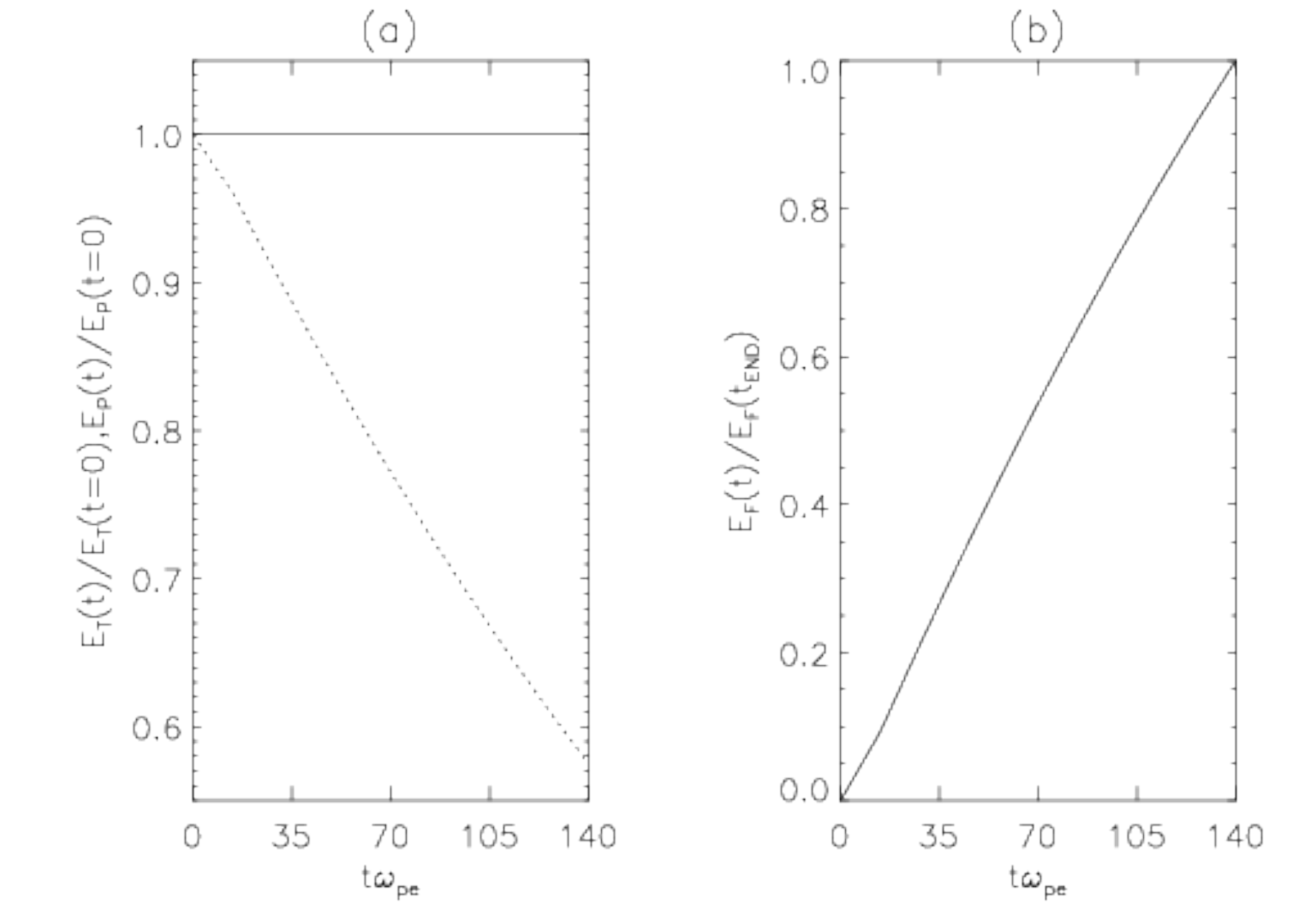}
\caption{Panel (a) solid and dashed curves respectively
are the total (particles plus EM fields) and only
particle energies, normalized on their initial values. 
Panel (b) shows
EM field energy, normalized on its final simulation time
value.}
\label{fig3}
\end{figure}

We used EPOCH, a
fully electromagnetic (EM), relativistic PIC code \cite{a15} for the simulation.
EPOCH is available for download from
\url{https://cfsa-pmw.warwick.ac.uk}.
The mass ratio is set
$m_p/m_e=1836.153$ for protons  and $m_i/m_e=85 \times 1836.153 $ for 
 rubidium vapor plasma.
The plasma temperature is set as twice
the ionization temperature of rubidium
i.e. $T=2\times 4.177 \mathrm{[eV]}\times 11604.505=96944$ K. 
Background plasma number density is set at $n_e=n_i=7 \times 10^{20}$ m$^{-3}$,
which lies within the AWAKE experiment range $n_e=n_i=10^{20}-10^{21}$ m$^{-3}$
The boundary conditions are periodic.
Choice of boundary conditions is not
essential because simulation domain is long enough,
such that the bunches never reach the boundary.

The simulations domain has
$n_x =4096$
grid cells in x-direction.
In y- and z- directions we have 24 grid cells in each direction.
In all directions, 
we fix grid size $\Delta$ as Debye length ($\lambda_D$) times appropriate factor ($10$), i.e.
$\Delta= 10 \lambda_D $.
Here $\lambda_D = v_{th,e}/\omega_{pe}$ denotes the Debye length
with 
$v_{th,e}=\sqrt{k_B T/m_e}$ being electron thermal
speed and $\omega_{pe}$ electron plasma frequency.
Here
the relevant spatial scale is electron inertial length,
 $c/\omega_{pe}$, which is
resolved with 25 grid points, i.e.
$(c/\omega_{pe})/\Delta=24.7 \approx 25$. This provides a good
resolution
as the total energy error never exceeds $\approx 0.004\%$.

At $t=0$, the driving proton and trailing electron  bunches have the number
densities as follows:
\begin{equation}
n_{D}(x)= A_D n_0  \exp\left[-\frac{(x-15.0 c/\omega_{pe})^{16}}
{(5.0 c/\omega_{pe})^{16}} \right]  
\label{e1},
\end{equation}
\begin{equation}
n_{T}(x)= A_T n_0  \exp\left[-\frac{(x-7.5 c/\omega_{pe})^{16}}
{(c/\omega_{pe})^{16}} \right]  
\label{e2},
\end{equation}
$A_D=100$ and $A_T=1$ are the bunch amplitudes in units of $n_0$.
We set 
$p_x=p_0=\gamma m_e 0.9999c$ kg m s$^{-1}$ 
(note that $p_x/(m_e c)=70.7$, i.e. $\gamma=70.7$),
which corresponds to an initial
energy of $E_0=\gamma m_e c^2=36.1$ MeV for electrons.
For protons the initial energy is $E_0=\gamma m_p c^2=66.3$ GeV.
In the simulation there are four  plasma species present:
background electrons and rubidium ions, plus
driving proton and trailing electron bunches.
In the numerical runs there are 
$50$ particles per cell for each of the four species.
i.e. total of $4096 \times 24 \times 24 \times 50= 117964800$
per each species.
The numerical takes about 16 hours on 128 cores using
Intel Xeon E5-2683V3 (Broadwell) processors
with 256GB of RAM and Mellanox ConnectX-4 EDR Infiniband Interconnect.

Fig.\ref{fig1} top row, panels (a-c), shows 
electric field x-component 
at different time instants
corresponding to 1/5th, half and the final simulations times.
We see that as the proton bunch moves in plasma, it generates a wake of
strength 
 $-2.5 \times 10^{12}$ V/m.
This is three orders magnitudes larger than that of AWAKE design, i.e.
of the order of $2 \times 10^{3}$ GV/m.
Commensurate plots in 
panels Fig.\ref{fig1}(d-f) show that 
trailing electron bunch remains always in the
region of nearly flat (constant) negative
electric field and hence is vigorously accelerated.
Also, Fig.\ref{fig1}(d-f) shows that the both bunches
stay intact by the end simulation time of $t \omega_{pe}=140$.
We see from panels Fig.\ref{fig1}(g-i)
the electric field 
x-component time evolution,  but now shown for the entire simulation domain
 at the same times. 
Note that x-coordinate is different in panels (a-f) 
because we use a window which follows the 
bunch with speed $v_b=0.9999c$.

There are two effects that work against an efficient electron acceleration:
(i) depletion of either driving laser pulse or proton bunch and
(ii) de-phasing of the trailing electron bunch from the negative electrostatic
plasma
wake. 
Only negative electric field can accelerate the electrons.
The positive one causes deceleration.
Comparing panels (a-c)  
to panels (d-f) in Fig.\ref{fig1} we see that trailing bunch is co-spatial with 
negative $E_x$. 
Thus driving bunch will be decelerating, because it gives off energy
to generate the wake,
while trailing bunch accelerating, because of sign of $E_x$.

In Fig.\ref{fig2} we plot 
background electron (a), rubidium (background plasma) ion (b), trailing electron bunch (c) and 
driving proton bunch (d) 
distribution functions, at different times in different colors:
open diamonds correspond to $t=0$, while blue and red curves to the 
half and the final simulations times, respectively. On
x-axis the momenta are quoted in the units of relevant species mass times
the speed of light i.e. $[m_e c]$, $[m_i c]$ or $[m_p c]$ as appropriate. 
At the top of panels (c) and (d), to guide the eye,  
the energies are stated in GeV with red numbers.
We gather from panel (a) that background electrons develop
broad peaks for positive momenta.
In panel (b) we see that background plasma rubidium  ions also
develop broad tails.
Panel (c) shows that by the end of simulation
 the trailing bunch gains energy to 60 GeV (red curve), 
 starting from initial 36.1 MeV.
Panel (d) shows that by the end of simulation
 the driving proton bunch loses energy from 66 GeV down to 3.6 
 GeV (red curve).
This demonstrates that trailing electron bunch acceleration is
due to deceleration of driving proton bunch. Similar conclusion 
follows from the dynamics of different kinds of energies, shown in Fig.\ref{fig3}.

In  Fig.\ref{fig3} panel (a) solid and dashed curves respectively
are the total (particles plus EM fields) and only
particle energies, normalized on their initial values. 
Panel (b) shows
EM field energy, normalized on its final simulation time
value. Because at $t=0$ all EM fields are zero, hence initial 
EM field energy cannot be used for normalization. 
The total normalized energy 
stays constant and is approximately unity.
Its maximal deviation from unity is 0.00004 i.e. $0.004 \%$
is due to numerical
heating and numerical dissipation (due to finite differencing).
The particle energy decreases by 42 percent.
The particle energy decreases because of deceleration of driving bunch
which then generates plasma wake -- the relativistic Langmuir waves,
which are then absorbed by the trailing bunch.
We see from panel (b) that
EM field energy normalized to its final simulation time
value increases, due  to the decrease of particle energy in panel (a),
so that the total energy stays constant within the error margin of $0.004 \%$.

To summarize, 3D particle in cell simulations were carried out to study 
proton-driven plasma wake-field
acceleration that uses a high-energy proton
bunch to drive a plasma wake-field for electron beam acceleration.
A new parameter regime was found
which generates essentially constant electric field
three orders magnitudes larger than that of AWAKE design, i.e.
of the order of $2 \times 10^{3}$ GV/m.
This is achieved in the extreme blowout regime, when number density
of the driving proton bunch exceeds plasma electron number density 100 times.
By the time of 
140 plasma periods trailing bunch gains energy to 60 GeV, 
 starting from initial 36.1 MeV.
The described in this work set up seems to provide
the most efficient electron acceleration.
Other runs (not shown here) such as: \\
(i) less dense proton beam (e.g. $A_D=10$ as opposed to $A_D=100$
considered here), \\
(ii) bunches localized in the transverse y- and z- directions,\\
(ii) bunches with higher $v_b$ (e.g. $v_b=0.999999c$ as opposed to
$v_b=0.9999c$ considered here),\\
show less efficient trailing electron bunch acceleration.

It is hoped that the exciting numerical simulation results
formulated in this work will find appropriate
experimental validation in facilities such as
AWAKE in the near future.
Indeed, the considered physical parameters are similar:
(i) AWAKE uses $3 \times 10^{11}$
protons per bunch and its
transverse dimensions are $200 \mu$m.
The latter corresponds to 0.996 $c/\omega_{pe}$ (for 
$n_e=n_i=7 \times 10^{20}$ m$^{-3}$).
Our transverse dimensions are similar 
$24 \Delta= 24 \times 10 \lambda_D$ that
correspond to 0.970 $c/\omega_{pe}$.
(ii) The proton bunch dimensions, based on Equation (1)
are approximately $10 \times 0.97 \times 0.97$ $(c/\omega_{pe})^3$.
Multiplying this by $100 \times 7 \times 10^{20}$
(as the driving proton bunch exceeds plasma electron number 
density 100 times) gives $5.33 \times 10^{12}$
protons per our bunch. This is an order of magnitude
larger that AWAKE protons per bunch, but reducing
transverse dimensions by $1/3$rd would bring it back
the feasible by AWAKE range.

Finally, it should be noted that using quasistatic approximation and with 
the wave-like ansatz, one can 
reproduce the analytical profile for corresponding proton beam. 
This profile should match
with the EPOCH solution. Such calculation would need to 
follow the steps given in Ref.\cite{bera16}
by replacing electron beam with the proton beam. 
This would also help to calculate the
transformer ratio analytically and to understand the 
underlying mechanism of proton-driven PWFA for
several beam densities and beam length, beam velocity.
Author would like to thank an anonymous referee for 
highlighting this very interesting point.

%{\bf Ethics statement.} Not applicable.

%{\bf Data accessibility statement.} This work does not have data.

%{\bf Competing interests statement.} The author has no competing interests.

%{\bf Author contributions.} Entire work was undertaken by author.

\begin{acknowledgments}
This research utilized Queen Mary University of London's (QMUL) 
MidPlus computational facilities,       
supported by QMUL Research-IT.
Author would like to thank two anonymous referees for 
useful suggestions, which improved this manuscript.
\end{acknowledgments}

%%%\bibliography{awake}

\begin{figure}[h!]
\includegraphics[width=3.5cm]{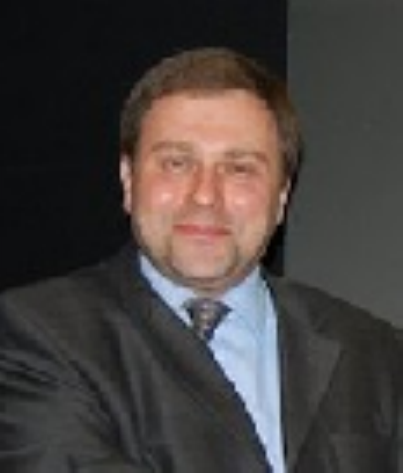}
%%\captionsetup{labelformat=empty}
\caption{David Tsiklauri was born in Tbilisi, Georgia in 1972. 
He received the MSci degree in Theoretical Physics 
from Tbilisi State University in 1994 
and Ph.D. in Physics degree from
the University of Cape Town in 1996.
He worked as Post Doc at University of Cape Town, 
Tbilisi State University, Iowa State University and University of Warwick. 
From 2003 to 2009, he was a Lecturer then Reader at 
University of Salford, United Kingdom. 
In 2009, he joined Queen Mary University of London, 
where he is currently a Senior Lecturer. 
His current research interests include novel 
particle accelerator concepts; plasma wake field 
acceleration; enhanced dissipation of MHD waves in 
inhomogeneous plasmas; collisionless magnetic reconnection; 
particle acceleration by dispersive Alfven waves and 
radio emission generation mechanisms by accelerated electrons.}
\label{dt_photo}
\end{figure}

\end{document}